# OmniDroneX: An LLM-Assisted Holistic Drone-as-a-Service Ecosystem

I-Ling Yen
University of Texas at Dallas
ilyen@utdallas.edu

Akeem Mohammed
University of Texas at Dallas
axm240136@utdallas.edu

Farokh Bastani
University of Texas at Dallas
Farokh.Bastani@utdallas.edu

San-Yih Hwang
National Sun Yat-sen University
syhwang@mis.nsysu.edu.tw

**Abstract.** Despite rapid advances in UAV technologies, current deployments remain limited due to several gaps in UAV systems research. To address these challenges, we propose OmniDroneX, a unified Drone-as-a-Service ecosystem, in which drones are transitioned from fixed-function platforms into dynamically composable entities that can be integrated with external infrastructures to offer "omni-capabilities." OmniDroneX bridges low-level physical primitives with high-level mission intent through a unified vendor-agnostic interface (libUAV) and a formal physical–service abstraction model (PT-SOA).

A core innovation is the diverse application of large language models (LLMs) across multiple layers of the OmniDroneX architecture. LLMs are used to assist in identifying and formalizing primitive device functions and abstract service definitions, supporting automated service composition and workflow generation, and enabling interactive, natural-language mission specification and refinement. OmniDroneX also incorporates important categories of composition techniques that are essential in dynamic UAV systems, including physical layer composition for drone capability augmentation, as well as spatiotemporal, functional, collaborative, exception-aware, and QoS-based service compositions. Collectively, these features allow OmniDroneX to serve as a foundation for scalable, resilient, and self-evolving UAV ecosystems operating in complex and dynamic environments.

**Keywords**. Drone-as-a-Service, unmanned aerial vehicles, large language models, service composition, Internet-of-Things, Cyber Physical Systems, digital twins, UAV marketplace.

## I. INTRODUCTION

Recent advances in unmanned aerial vehicles (UAVs) have enabled a diverse range of applications, including infrastructure inspection, environmental monitoring, precision agriculture, logistics, public safety, disaster response, and smart-city services. As drone technologies mature, future applications are expected to involve increasingly complex missions requiring coordination among heterogeneous drones, diverse sensors and IoT devices, sensor networks, multi-modal communication infrastructures, edge/cloud computing resources, and specialized software services. Yet, despite the extensive body of UAV research, current drone ecosystems remain fragmented and expose several research and engineering gaps.

First, drones from different manufacturers expose incompatible interfaces, support varying sensing and communication capabilities, and rely on proprietary software stacks. Mission developers are often forced to manually integrate heterogeneous APIs and infrastructure components, resulting in significant development effort, limited interoperability, and high system maintenance costs.

To address this gap, we propose libUAV, the foundational layers in OmniDroneX, designed to provide "one interface to rule them all" (the slogan of Apache iCloud). We leverage large language models (LLMs) with guided prompts to automatically analyze API/SDK documentation from diverse drone vendors and generate a standardized set of primitives. Regardless of mission domain, fundamental operations such as route planning, autonomous navigation, payload manage-ment, power management, and sensor control are essential for mission execution. From this unified abstraction layer, libUAV automatically translates standardized UAV operations into vendor-specific API calls, allowing mission developers to focus on mission logic rather than platform-specific implementation details. This abstraction promotes portability, greatly reduces development effort, enables interoperability across heterogenous UAV hardware and platforms, alleviates hardware and vendor lock-in, and serves as a foundational building block for higher-level service composition and mission orchestration for scalable UAV applications.

The second and third gaps are interrelated: the second concerns the absence of well-defined UAV service models, and the third concerns deficiency in UAV service composition techniques built upon such formal models.

Some Drone-as-a-Service (DaaS) works focus on individual application domains such as delivery [1] [2] [3] or communication [4]. Broader paradigms, such as those proposed in [1] [3] simply suggest abstracting drone functionalities as services. [5] focuses on DaaS, identifies several categories of fundamental services, and discusses issues in service selection and composition, but remains a survey without concrete DaaS solutions. [6] identifies a set of primitive services for drones, sensors and compute units, showing through case studies that applications can be implemented with a small number of lines of code when building upon these services. However, the applications it focuses on involve only single drones and are limited to sensor data collection type of applications. Also, the paper has limited discussion of service concepts.

The concept of "dockerizing" drones (their on board computers) introduced in [7] is interesting and practical as it makes built-in vendor software stacks replaceable, facilitating the realization of DaaS and vendor-agnosticity.

In [4] [8] [9], the UAV paradigm is generalized and integrated with IoT systems, but these works treat UAVs only as the communication service providers that forward IoT data to IoT platforms, edge, and/or cloud. [9] [10] explicitly integrate UAVs with edge and cloud, which offer storage and analytics computation services. In general, neither these works nor other service abstraction approaches provide a formal model for UAV services or address how UAV services can be composed to fulfill user-specified missions.

Some UAV research considers multi-drone compositions, but nearly all these compositions are spatiotemporal based [1] [2] [11] [12] [13], such as selecting drones along a computed route or partitioning a large region and assigning the same tasks in different subregions to different drones. Some of them further incorporate quality of service (QoS) factors, such as efficiency and energy consumption. Some others consider composition of UAVs to form ad hoc networks [4] [8], which remains spatial-(coverage-) based composition. While spatiotemporal-based UAV composition is important, functional and collaborative composition paradigms are equally essential to fully realize UAV system potential.

In OmniDroneX, we address the service modeling gap by adapting our earlier works in PT-SOA [14] [15] [16], which defines service semantics for cyber physical systems (CPS) and IoT systems, to UAV-centric systems. Note that software services and services provided by physical entities have fundamental differences. In a simple example, a software service can be used in multiple workflows, whereas physical services are constrained by the availability of the underlying entities. Existing works in drone service composition address availability through scheduling and resource allocation, but ignore the service modeling perspective, which can complicate the semantics of service composition. If a physical entity provides multiple services, should availability be considered at the physical entity level or at the level of its services. Existing composition works only address resource allocation [1] [2] [11] [12] [13] [17] rather than service composition, effectively treating physical entities as resources rather than as services, despite conceptually referring to the DaaS abstraction.

UAV service composition should extend beyond drones and spatiotemporal coordination to include other physical entities and other forms of composition. An important concept that is overlooked in the literature is that the services a drone can offer are not fixed, but can effectively be unbounded via physical augmentation [14] [15] [16]. For example, a drone can carry different sensors to enable distinct sensing capabilities. Similarly, mounting additional communication modules, such as WiFi, cellular, or radio devices, to a drone expands its communication modality. These forms of physical composition significantly expand UAV's capabilities, but they also introduce complexity in composition reasoning. For example, the flight of the drone shifts the location of the loaded sensors, and the sensors will no longer perform designated sensing in its original location, but at where the drone is. Additional semantics should be incorporated in service IOPE definitions. Software can also be loaded onto drones to perform different flight control, sensor data integration, analytics, etc. at the location of the drone. Moreover, vehicles can further enrich the UAV ecosystem by transporting drones to the last mile to reduce its flight time and battery consumption [18] [19]. Thus, UAV service composition should adopt an expanded CPS/IOT perspective.

Another largely overlooked direction is collaboration-based composition. For instance, while a single drone has limited payload capacity, multiple drones can collaboratively transport a heavy object for delivery or disposal [20]. In bucket-brigade delivery scenarios, drones may sequentially transfer payloads under coordination protocols that ensure robustness and fault tolerance [21]. As another example, in drone-as-a-guide applications, the drone-guide cannot have its independent navigation planning, but must closely synchronize with the group to accomplish the mission. Such collaborative behaviors cannot be adequately captured by conventional composition paradigms, particularly in automated service composition, as they extend beyond independent service IOPE specifications and require richer collaboration semantics [22].

At the core of OmniDroneX is a composition agent. It treats drones as first-class entities within a broader cyber physical ecosystem. Our DaaS framework allows drones, sensors, and software services to be composed at both physical and service levels to achieve augmented capabilities. External resources such as communication infrastructures, charging stations, mobility platforms, sensor networks, and edge/cloud computing services are incorporated in OmniDroneX and can be discovered and composed dynamically according to mission requirements. This enables the construction of complex mission plans from a rich set of cyber physical entities and services.

Based on the extended UAV service model, the composition agent leverages LLMs and AI planning techniques for automated service composition. OmniDroneX can translate user intents specified in natural language into PDDL representations and automatically generates optimized mission plans. LLMs are also used to assist in defining UAV services based on various web resources and natural language descriptions of the physical entities and their services. New composition paradigms are introduced to support both physical entity composition and collaborative composition.

Orthogonal to DaaS, which focuses on services and composition, a marketplace paradigm enables the sharing of UAV resources and services among diverse stakeholders. It matches user requests with available resources and services in a supply-demand framework similar to platforms like Uber or Airbnb. While marketplace technologies are mature, existing platforms primarily handle independent items or simple services, and they do not directly support composite services. On the other hand, manufacturing-as-a-service marketplaces, such as Xometry and Fictiv, demonstrate how composite services can be matched, scheduled, and executed to fulfill user requests. UAV missions often involve similar complexities, yet marketplace platforms specifically for composite UAV services remain underexplored, representing a significant gap.

OmniDroneX addresses this gap by providing a marketplace layer built on top of the DaaS model, with composition agent from the lower layer as its core. This marketplace does not operate in isolation but interacts with existing sensor networks, vehicle networks, and edge/cloud computing platforms, forming a complete ecosystem. OmniDroneX offers a practical and scalable marketplace by adapting the design principles in manufacturing-as-a-service marketplaces to the unique needs of UAV mission.

A final gap concerns architectural designs of existing DaaS frameworks. Existing layered architectures are mostly limited in scope. [5] presents a very basic architecture. [23] focuses mainly on individual drone capabilities, rather than architectural design.

Those that integrate IoT [4] [9] or cloud services [10] consider drone layer simplistically, limit drones' potential in the system.

OmniDroneX addresses this gap by introducing a multi-layered architecture that revolves around the lifecycle of UAV applications, from physical entities, to service abstraction, composition, validation, and execution, complemented by an orthogonal marketplace. OmniDroneX also adopts a broader CPS/IoT systems perspective in which drones, sensors, software services, and various external infrastructures are treated as composable entities within a unified ecosystem. Moreover, OmniDroneX emphasizes the linkage between layers, offering a clear path for incremental layer by layer development.

The contributions of this paper are threefold. First, we identify the key research and engineering gaps in existing service-oriented UAV systems and frameworks. Second, we present a layered architecture for realizing an integrated DaaS ecosystem. In its name, "Omni" signals broad coverage across the full spectrum of UAV development lifecycle; "Drone" anchors the framework in its technological core; while "X" denotes extensibility, highlighting its capacity to compose, evolve, and encompass the broad CPS/IoT ecosystems. OmniDroneX is designed to bridge current gaps and enable a unified UAV ecosystem. Third, we outline a research agenda highlighting the challenges and opportunities associated with building large-scale drone service ecosystems.

The remainder of this paper is organized as follows. Section II reviews related work. Section III presents the OmniDroneX architecture and its design objectives. Section IV discusses open research challenges and future directions, followed by concluding remarks in Section V.

# II. RELATED WORK

Most of the research works in UAV focus on specific applications. Drones have been widely used for monitoring applications, such as monitoring the air quality, waste management, marine monitoring [24] and inspection for critical infrastructures such as power lines [25] [26], hydraulic structures [27] [28], bridges [29] [30], rail tracks [31], etc., for civil structures such as buildings [32] [33], industrial facilities [34], etc., for precision agriculture [35] [36], etc. Autonomous delivery systems are another major application area of UAV. [37] simply compares UAVs and land robots for last mile delivery. [3] considers the delivery of COVID-19 self-test kits to potential patients with optimized drone path. [2] [18] consider the problem of assigning drones to different segments of a delivery route to accomplish cooperative delivery, a typical spatiotemporal composition approach. There are also research works focusing on forming ad hoc networks using drones' communication capabilities [4] [8] [9]. These works abstract drones as services but only use them for transferring IoT data. [38] summarizes what UAVs can do in disastrous situations, including disaster detection, assessment, and response and recovery. Though considering making use of drones capabilities in monitoring, communication, and delivery during disasters, the paper placed a heavier focus on providing urban communication infrastructure by UAV fleets to facilitate disaster management.

## A. DaaS Abstractions and Frameworks

The general Drone-as-a-Service (DaaS) paradigm has been explored with the central idea of abstracting UAV systems into service-oriented infrastructures. [5] surveys many aspects of drones and DaaS. It introduces drone properties and categories of various applications in the literature. Then it considers the DaaS architecture: stating that a typical DaaS system architecture comprises three layers: drone, computing, and user (including drone owners or drone users), which has a focus on UAV marketplace. The paper does consider identifying basic services, including sensing functions, inspection functions, delivery functions, entertainment functions, videography functions, communication functions. Generally, services are identified to be composed into applications. Some of the service categories identified in this paper are reasonable, but others are almost at the applications level. For example, entertainment and videography applications in fact make use of the sensing capabilities of the drones. They differ from general monitoring applications at a high level, regarding what to sense and how to deliver the results. The research challenges discussed in the paper lack clear categorization, including DaaS communication and data management, environmental uncertainty prediction, DaaS cost estimation, DaaS safety assurance, DaaS scheduling, DaaS energy consumption optimization, DaaS selection and composition, DaaS crowdsensing/crowdsourcing, DaaS cybersecurity, human–drone interaction (focusing on the interaction with the DaaS platform). Service selection and composition is indeed a core part of DaaS. However, DaaS scheduling, DaaS cost estimation, DaaS energy consumption optimization, QoS and uncertainty issues, environmental uncertainty prediction, are part of the selection and composition issue. QoS, cost estimation, energy consumption, etc. are the QoS aspects of the composition problem. Uncertainty is an additional considerations during composition to ensure mission success. Scheduling is a core technique for spatiotemporal composition. On the other hand, data management, safety, cybersecurity, human–drone interaction are DaaS platform design issues for high assurance.

AeroDaaS [6] considers a service-oriented UAV framework. It focuses on identifying a set of APIs (as services) to support easy, hardware-agnostic implementation of UAV applications. The set of primitive operations include querying available sensors, robots, and compute devices, retrieving their properties, collecting data, offloading analytics to computing resources, and generating navigation commands based on the analysis results. The paper shows that, given their API set, fewer than 50 lines of code are needed to implement a task. This work solves a real programmability problem, and does offer useful hardware-agnostic APIs. However, the tasks considered are primitive and homogeneous, each involving a single-drone setup with matching sensors to obtain the desired sensor data. The framework does not consider multi-drone collaboration, and the only form of composition is the offloading of analytics to the cloud. Moreover, although the paper is titled as "DaaS," but the framework lacks service concept or service composition.

[7] proposes a novel concept of dockerizing drones' on board computers so that they can be seamlessly integrated with the

cloud to perform flight control and other tasks. This allows the vendor-specific software stacks to be replaceable and adaptable, making it easy to achieve vendor-agnostic development of the high-level systems and solving the problem when existing software stack has deficiency in fulfilling some missions. The paper also considers the networking issues. Client and drone can communicate using REST APIs. Port forwarding and other networking mechanisms are utilized to facilitate the communication between users and drones when IP address changes.

[39] presents the design of a Drone-as-a-Service (DaaS) system. It identifies the relevant stakeholders, including mission definer, mission supervisor, mission viewer, pilots, payload manager, system manager, and logistics personnel. It also specifies the functional and non-functional requirements of the system, together with a detailed microservice-oriented architecture. In the case study validation, a set of mission attributes are defined, including: sensors/actuators required (e.g., thermal camera, toxic analysis sensors, video streaming), the operation zone (geographic location and radius where the mission will take place), and the landing point (designated location for drone landing after mission completion, flexible depending on mission progress). Based on these attributes, the system selects a drone, assigns a pilot and logistics personnel, and automatically generates flight plans for the selected drone, from takeoff to landing. The missions considered in this work are simple and restricted to single-drone operations. Matchmaking (resource allocation) is basic, relying only on the availability and proximity of drones and sensors. Moreover, the work has limited relevance to services. It does not define service abstractions for drones or other devices, considers no entities beyond drones and sensors, and does not address service composition. Thus, the contribution is essentially a system design rather than a true service-oriented paradigm.

While all these works consider "X-as-a-Service" (X=Drone or IoT) or service-oriented, none of them consider service modeling specifically for UAVs or consider extensively the service composition issues, which are core issues in DaaS. Most of the existing works treats UAV or IOT services the same as other services in the literature. In contrast, [14] [15] [16] not only treats IoT devices as services, but also proposes a formalized PT-SOA model. They also discuss the differences between CPS/IoT services and traditional software services. An IoT service involves a provider (such as drone) and the services. A service can be offered by multiple physical entities which may be heterogeneous, with potential physical constraints, etc. Shall physical-entity (PE) specific features and constraints be specified via services? No! So, the paper suggests the separation of PE and their services while addressing their interdependencies via the PT-SOA model. Features and constraints on PEs can be specified with the PE, and IOPE for services are specified with the services. In [14], a formal service-oriented model using ontology-based definitions is presented, which enables specification, verification, and automated reasoning about CPS/IoT services. [15] highlights the deviation of physical services from software services, stressing how IoT services must incorporate context-awareness, physical constraints, and dynamic availability. [16] focuses on identifying the gap between software service modeling and IoT service modeling, proposing extended semantics to represent physical entities and their service representations.

### B. UAV Service Marketplace

In [40], Aerialis is proposed as a marketplace for aerial computing. The paper introduces an 'Interest Map' interface to represent UAV tasks, where each task is characterized by its location, estimated duration, task executable, and a monetary bid reflecting the user-assigned reward for task completion. Though the concept is interesting, using monetary values as a proxy for task priority offers limited advantage over directly specifying priority levels, which many existing systems already support. If monetary values are intended to model actual mission cost, user-specified bids are insufficient and accurate cost estimation in UAV missions must account for flight distance, battery consumption, resource availability, sensor usage, and infrastructure fees, none of which can be reliably self-assessed by the user. Furthermore, a meaningful marketplace mechanism requires negotiation between requesters and service providers, and potentially among providers themselves when composite missions are involved. Aerialis does not address such negotiation, functioning in practice as a priority-based task dispatcher rather than a true marketplace. The matchmaking mechanism relies on pre-computed spatial clustering based on drone locations, which enables efficient assignment of independent tasks within geographic regions but does not support dynamic service recomposition or functional coordination across heterogeneous resources. Consequently, Aerialis addresses location-based drone-to-task matching but falls short of the composite service orchestration and dynamic mission coordination required for complex UAV operations.

[5] considers DaaS like a UAV marketplace, and the system includes three stakeholders, users, drones, and compute. But the discussion in this aspect is limited.

Community Drones [41] proposes a shared UAV infrastructure where drones are accessed as pooled resources. The system consider 3 stakeholder roles: clients who propose drone related missions, the infrastructure operator and mission operators and 4 components: flight planning, weather service, risk assessment, and scheduling. In the design, the two operator roles both focus on execution and are also the providers. The inclusion of a regulatory framework is a highlight of this paper, and is essentially the 5th component in the system design. Overall, the main focus of the paper is on automated mission planning, which is indeed a core part in any UAV marketplace. Specifically, flight path planning is done in three steps: map analysis, scanning pattern generation, and flight path computation. When generating scanning patterns, risk assessment and regulatory framework are taken into account to avoid risky or nonpermissive regions. For generating the optimal flight path, well established A* or Rapidly Exploring Random Trees (RRTs) search algorithms are suggested. Scheduling is a another core part of mission planning and in this paper, it considers optimization for missions of multiple clients. The derived schedule should consider task priorities, safety, data quality, cost efficiency, and sustainability. Schedules are derived using a Variable Neighborhood Search with localized improvements through Variable Neighborhood Descent.

UAV marketplace platforms differ from familiar platforms such as Uber or Airbnb where each trade only involves independent items. UAV marketplace must support the collaborative use of multiple drones and providers to accomplish complex missions. Thus, composition should be the core of the UAV marketplace and negotiation between the requester and providers and among the providers are essential. Some of the issues are addressed in Manufacturing as a service platforms such as Xometry and Fictiv. However, the situation dynamics and responsiveness requirement in critical UAV missions remains a unique challenge. UAV marketplaces must orchestrate and adapts instantly to dynamic mission conditions, a requirement that has not been fully addressed by existing marketplace designs. [40] does not consider dynamic composition, but only consider location based allocation of drone fleets for each independent task. [41] handles dynamic changes, but only focuses on flight path planning and scheduling, does not handle functional compositions or compositions with external infrastructures.

### C. UAV Service Composition

Several works formulate mission planning as a QoS-aware resource allocation problem within the DaaS paradigm with varying emphases on QoS constraints, application domains, and algorithmic mechanisms. Rather than functional service composition, these approaches focus on matching mission workflows of UAV services to available drone resources subject to specific constraints.

[9] presents a layered service-oriented architecture that conceptualizes drones as first-class IoT services. The framework introduces a service registry to track available drones, their capabilities, and their QoS properties (energy, latency, cost, etc.). A reporting layer monitors the execution of drones and record the QoS related status, such as latency, power consumption, utilization, etc. The focus of the paper is on service selection, which is based on QoS requirements of the tasks and demands by the providers.

[11] addresses energy as the central QoS dimension, proposing in-flight recharging and energy-aware allocation. It employs enhanced A* search for optimal service composition, along with priority-based and fairness-based energy sharing strategies. [1] models the composition problem under unpredictable conditions such as weather or battery variability. It introduces a spatiotemporal composer and a predicative composer. The spatiotemporal composer uses a rule-based approach to allocate drones to unit tasks. More specifically, it matches drones' spatiotemporal states, including their locations, battery conditions, payload capacity, and availability window, against the requirements of mission tasks to determine the assignment. The spatiotemporal composer uses machine learning classifiers trained on spatiotemporal features to predict drones' suitability for mission tasks under uncertainty.

Unlike earlier works that consider centralized resource scheduling, [13] introduces a spatiotemporal-aware decentralized service discovery framework. It employs a two-phase affinity propagation mechanism to elect leaders who serve as registries for their clusters and perform locally centralized service discovery in the clusters. Leader election is determined by the "suitability" score (suitability to be a leader). Specifically, suitability is defined as a weighted sum of spatial centrality (a centrally positioned node in the cluster is preferred), mobility alignment (moving in similar directions as other nodes is preferred), remaining storage/computation resources, lower workload (fewer important services hosted), and remaining battery power. When a drone request involves a certain cluster, the leader performs service discovery to identify the drones in the cluster based on the composite QoS scores of the candidates. The framework also considers resilience and introduces a priority chain mechanism for replication and failover: a leader is backed up by a ranked chain of drones ordered by their suitability scores. This algorithm achieves lower service registration and discovery latency and faster failure recovery compared to centralized cluster-based solution. However, it has a split-allegiance problem. This occurs when a boundary node (e.g., node B) lies between two leaders (A and C) that are not direct neighbors. While leader A may include B in its cluster, node B may consider C as a more suitable leader. Such conflicting preferences can lead to unstable cluster membership. More importantly, the paper does not discuss how missions are decomposed into subtasks or how those subtasks are routed to the appropriate cluster leaders.

Besides the general drone to task assignment problem, some works focus on a domain-specific application for resource allocation. [12] consider oil pipeline surveillance. It formulates the composition problem as a mathematical optimization problem, balancing coverage efficiency and surveillance cost. Different group sizes and base locations are evaluated to find the most effective allocation of drones. [2] emphasizes spatial QoS constraints, optimizing delivery missions by minimizing travel distance and latency. Its algorithmic contribution lies in knowledge graph–based composition with metapath embedding for proximity-aware service selection. Collectively, these works contribute concrete algorithmic mechanisms for resource allocation (service discovery is in fact an equivalent problem). They also highlight diverse QoS dimensions such as energy, uncertainty, proximity, and spatiotemporal awareness, and explore both general frameworks and domain-specific applications. However, they remain limited to resource allocation rather than functional service composition, with QoS constraints treated in isolation rather than integrated holistically.

Another form of composition considered in the literature is collaborative composition, which extends beyond traditional function composition or QoS-based service composition. Unlike traditional composition, collaborative composition requires close synchronization among service providers during service execution. Some works in swarm robotics illustrate the foundations of this paradigm. [20] demonstrates how distributed robots can autonomously build and maintain formations even under failures, where the system always can stabilize to a valid structure even after disturbances. It is highly relevant for drone applications such as disaster response through debris removal. Complementing this, [21] addresses the collective search, collection, and delivery of resources. By relying on simple local rules, swarms can self-stabilize and maintain efficiency even when individual robots fail. This approach is also directly applicable to drone fleets tasked with object transportation, where multiple drones must coordinate to locate, lift, and deliver items. Together, these works highlight the self-stabilizing paradigm for collaborative composition, enabling drones to function as cohesive, fault-tolerant collectives capable of executing complex missions. By embedding collaborative composition into DaaS, drone services can evolve from loosely coupled task execution toward cohesive, fault-tolerant collectives capable of tackling complex missions that demand both adaptability and tight coordination.

In OmniDroneX, we adopt and unify these existing composition techniques and introduce new ones to enable larger-scale and more complex UAV missions.

## III. System Architecture

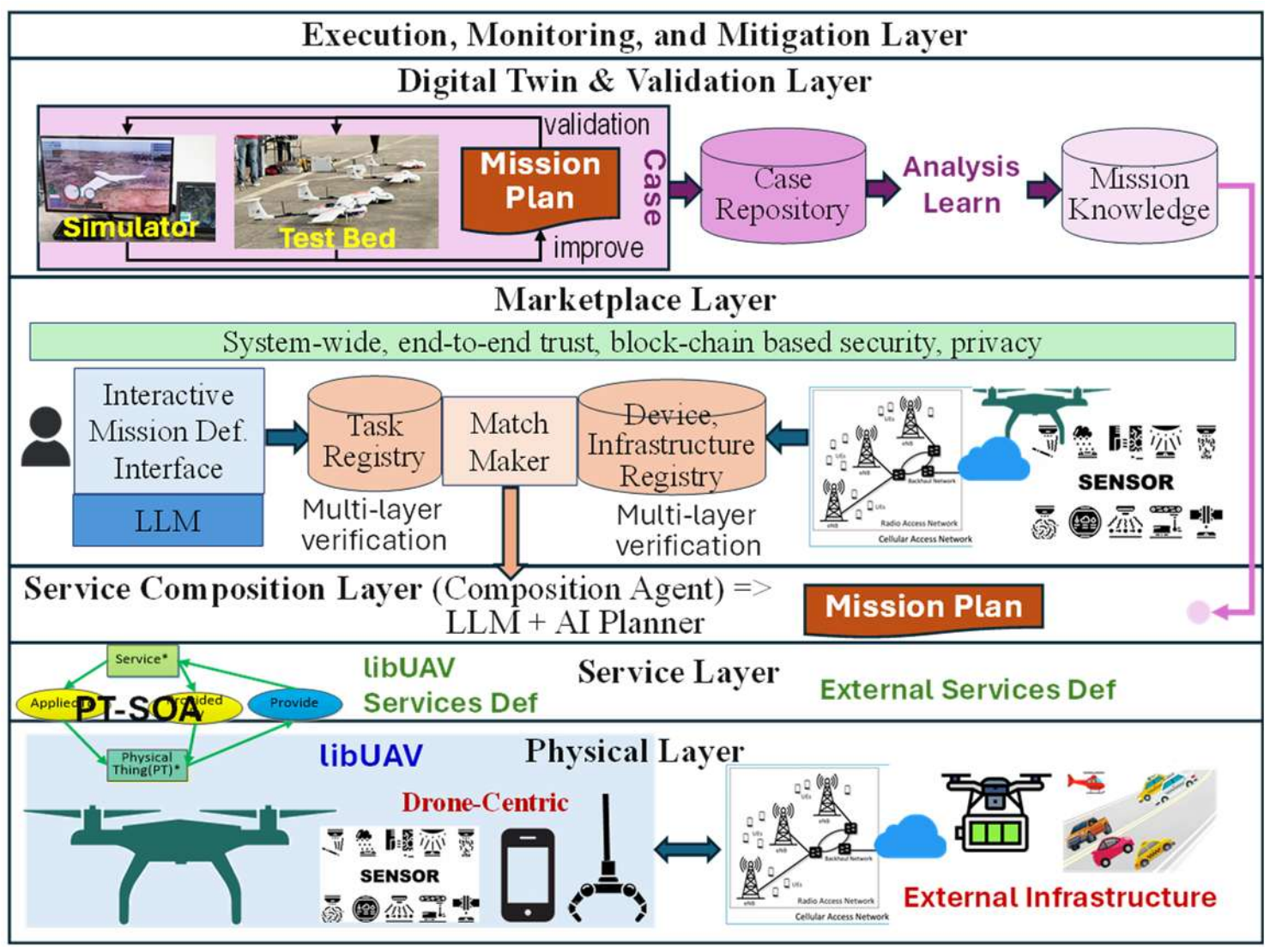

Fig. 1. OmniDroneX system architecture.

In OmniDroneX, we design a layered architecture that revolves around the lifecycle of UAV systems. As shown in Fig. 1, the architecture consists of six layers, each providing distinct yet comprehensive functionalities to support the layers above it and enable a full-fledged Drone-as-a-Service ecosystem. The physical layer consists of drones and other devices as well as physical infrastructures to support the encompassing capabilities of the ecosystem. The service layer builds on the physical layer to provide abstractions for higher level vendor-agnostic development. UAV missions are composed from services by the composition agent in the composition layer. The UAV marketplace layer builds on the composition layer and provides intelligent support to assist users in specifying their missions. The architecture continues into simulation and testbed as well as execution, continuous monitoring, and mitigation layers to address the complete UAV ecosystem lifecycle. Several important and novel design concepts are highlighted below.

(1) At the foundation layers, OmniDroneX provides unified interfaces in libUAV to enhance interoperability and offer the notion of one interface to rule them all. The physical layer consists of physical entities and infrastructures and exposes primitive functions provided by these entities. Services in the service layer are defined on top of these physical functions. libUAV spans these two layers to support vendor-agnostic high level composition for user defined missions. To support automated service composition, we extend the formal PT-SOA model for physical entities and service specifications.

(2) Automated composition is one of the major focuses of OmniDroneX. It considers physical device composition, where non-native sensors and other devices can be mounted on a drone to augment its capabilities. In the composition layer, we categorize service composition techniques that are practical to the UAV application domain and provide all types of automated service compositions, not limited to existing spatiotemporal scheduling algorithms or QoS-based composition methods.

(3) The OmniDroneX ecosystem considers the integration of drone platforms with other infrastructures available in the smart world environment, including sensor networks, battery supply infrastructures, multi-modal communication infrastructures, battery supply infrastructures, mobility platforms, and edge/cloud computing, storage, and software services, to support large-scale complex missions and further enhance the comprehensiveness of the ecosystem.

(4) A marketplace platform is introduced in OmniDroneX to assists user specifying mission objectives using high-level natural language descriptions and LLM-based interactive refinement. OmniDroneX then automatically translates the final mission specification into executable service workflows to fulfill the mission requirements.

### A. The First Layer: The Physical Layer

The physical layer consists of various hardware devices, including drones, sensors, payload modules, communication devices, etc. We consider three categories of physical entities: (1) the *drones* themselves. (2) *drone-centric devices* , including devices that can be mounted on drones to enhance their capabilities. Examples include sensors, communication devices, payload devices, and computing devices. For example, a smartphone can be mounted to a drone and the drone will be enhanced with additional sensing, cellular communication, and computing capabilities. (3) *Externalinfrastructures* that can be integrated into UAV missions. At the physical layer, these infrastructures are considered as physical units, while their integration with drones is addressed at the service composition layer. Examples include (3a) sensor networks, (3b) communication infrastructures such as cellular networks and radio networks, (3c) edge and cloud computing infrastructures where drone can offload computing intensive analytics, (3d) battery supply infrastructures, such as battery swap stations, (3e) mobility infrastructures that can transport drones closer to mission locations to reduce their flight time and battery consumption, and (3f) knowledge infrastructures, which can provide information to assist UAV operations, for example, weather information obtained from web resources can be used during flight planning. These infrastructures typically expose APIs as services that can be invoked by UAV systems and can be viewed as infrastructure providers for drone missions, although they differ from conventional cloud Infrastructure-as-a-Service platforms.

A key concept of OmniDroneX is the **physical layer composition** for *capability augmentation*. The concept is elaborated in Fig. 2. Many existing UAV models assume a predefined hardware and software stack. Drone's specifications are frequently defined by a fixed set of capabilities. While some capabilities are inherently fixed (due to the built-in configurations), such as payload capacity, drone size, built in communication capability, etc., many others can be dynamically added or modified through physical augmentation. For example, mounting a life-detection sensor enables a drone to perform victim detection in emergency response missions. Similarly, a drone that natively supports WiFi communication can obtain cellular communication capabilities by carrying a smartphone. Therefore, drone capabilities should be viewed as dynamically composable at the physical layer rather than being permanently tied to a fixed hardware/software configuration. We consider this as physical layer composition. Correspondingly, how to *automatically infer the new capabilities for the composed physical object from the capabilities of the individual physical objects* should be investigated.

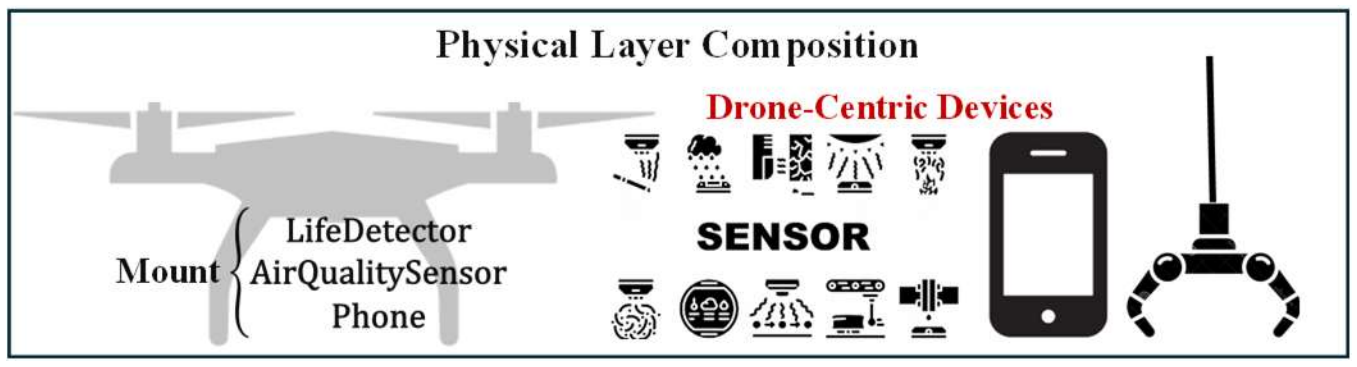

Fig. 2. Physical layer composition: Capability augmentation.

A second key feature in the physical layer of OmniDroneX is the development of **libUAV**, which follows a philosophy similar to Apache libCloud's "one interface to rule them all.". Current drone ecosystems are typically built as closed, vendor-specific platforms. In these vertically integrated designs, mission planning, navigation, sensing, communication, and payload management are all tightly bound to the vendor's hardware and software stack. This siloed approach makes drones from different vendors difficult to interoperate and increases the complexity of building large-scale UAV systems that involve multiple drones, diverse sensors, infrastructures, and software components.

The focus of existing UAV frameworks is to abstract drones as services without concerning the interoperability issues. Main effort in dealing with interoperability is by *STANAG 4586* [42], which standardize interaction protocols and navigation control functions. We plan to incorporate these functions in libUAV. However, STANAG only involves drones and focuses on the lowest level of functions and interactions.

We will use LLMs to analyze SDK documentation from different drone vendors and determine the additional primitive functions for drones, including flight-control operations, telemetry and status primitives. Equally importantly, the primitive functions for drone-centric devices should be defined. Some functions are common to all devices, such as device initialization, configuration, and reset. Other functions are device specific, such as connect(), send(), receive() for communication devices, start_video_streaming(), end_streaming() for cameras, pickup() and dropoff() for payload modules, etc. One important pair of functions for drone-centric devices are **mount**(drone, device) and **unmount**(drone, device), which are critical for physical composition.

Several research and engineering tasks are required to realize this layer. First is the development of libUAV. LLMs will be used to categorize hardware devices and identify their common primitive operations. Note that in physical layer, we consider primitive functions rather than services. To realize interoperability, functions in libUAV need to be translated into the vendor-specific functions and LLMs will be used to assist with the generation of such translation code based on vendor specific SDK documents. Second, physical entities should be formally defined using the PT-SOA model to facilitate the upper level service definitions and compositions. LLMs will also be used to generate physical-entity specifications from existing documentation for drones and other devices.

## B. The Second Layer: The Service Layer

The second layer of OmniDroneX is the Service Layer. While the physical layer defines primitive functions provided by individual physical entities, the service layer defines reusable capabilities that can be offered by a single drone together with its mounted devices, or by an external infrastructure integrated into the drone ecosystem. Services are built on top of primitive functions and may additionally involve planning algorithms, data processing, analytics functions, or decision-making logic. The purpose of this layer is to elevate low-level device operations into meaningful services that can be invoked and composed into more complex workflows to fulfill user mission requests.

One important category of services is flight and navigation services. While the physical layer provides primitive flight-control functions such as movement and hovering, the service layer provides navigation capabilities that incorporate planning and reasoning. For example, a *route-planning* service generates a sequence of waypoints between a source and destination while considering flight restrictions, environmental constraints, and quality-of-service requirements. A *flight-to* service further combines route planning with waypoint execution to autonomously navigate a drone to a target location. More advanced navigation services *flight-to-object* may support flight toward an imprecisely specified target, such as a detected object, or perform precision landing on a designated target using visual and inertial guidance. Such services typically rely on additional perception services for target identification and localization.

Payload-related services represent another important class of drone services. Rather than exposing low-level gripping or releasing functions, the service layer provides higher-level capabilities such as object pickup and payload delivery. For example, a *pickup-object* service may involve autonomous navigation to a specified location, object identification, positioning, and physical retrieval of the target object. Similarly, a *dropoff-object* service may deliver an object to a specified destination or onto a designated target such as a building rooftop or a moving vehicle. Other payload services include $regional\text{-}payload\text{-}release$ service, where liquid or gaseous materials are dispensed gradually over an area, and $payload\text{-}deployment$ service that activate specialized payload modules (such as a bomb) before release.

Services can also be provided by mounted sensors leveraging the mobility of the drone. These services transform raw sensing functions into meaningful sensing operations. For example, the $sensing(location, duration)$ service may continuously collect observations at a specified location for a given duration, requiring the drone to maintain a stable hover while sampling the environment. Alternatively, $sensing(region, duration)$ may be performed over an entire region, requiring navigation functionality and data collection. The specific sensing modalities depend on the mounted sensors and may include environmental monitoring, radiation sensing, gas detection, or other forms of data acquisition.

Camera-equipped drones can further provide perception-oriented services. A $target\text{-}detection$ service identifies objects of interest within captured imagery and may be used to locate landing targets, vehicles, infrastructure components, or individuals. Building upon this capability, a $tracking(object)$ service continuously monitors the movement of a detected target object over time. Another related service is $following(object)$, in which a drone autonomously navigates by continuously tracking a linear object or feature such as a river, road, power line, or pipeline.

Communication devices mounted on drones provide communication services that abstract the underlying communication technologies. A drone may establish communication with another drone or with a ground-based communication device via $connect$ service. Such services must consider communication range, link quality, and the selection of appropriate communication technologies built-in or mounted to the drone. Beyond connection establishment, common $send$, $receive$, and $relay$ services are offered. Relay services are particularly important because they enable a drone to function as an intermediate communication node without necessarily participating in application-level communication itself.

In addition to services provided by drones and their mounted devices, the service layer also considers services provided by external infrastructures. These infrastructures are represented as physical entities in the physical layer and expose capabilities through service interfaces that can be invoked by drones and other system components. One example is the **battery supply infrastructure**. Although batteries are physically attached to drones, battery replenishment generally requires external support. Energy infrastructures may therefore provide $battery\text{-}replacement(location)$ service that allow drones to replenish energy resources at designated locations. Such services may be realized through charging stations, battery swapping facilities, mobile charging vehicles, or other facilities. **Mobility infrastructures** provide transportation services that complement aerial mobility. A $ride(src, dest)$ service allows a drone to request transportation between two locations using an external transporter. The drone first travels to a rendezvous location $src$, is transported by the mobility provider, and is subsequently released at the destination $dest$. Transporters may include dedicated drone trucks, conventional ground vehicles, watercraft, aircraft, or even commercial transportation services. Such mobility services can significantly extend operational range while conserving battery resources. **Sensor-network infrastructures** can provide sensing-related services independent of the drone itself. These services may collect data from distributed sensors deployed throughout an environment or relay collected observations to designated destinations. Similarly, **communication infrastructures** such as cellular networks provide communication services that allow drones to establish connectivity with infrastructure nodes. In many cases, a drone only needs to connect to a nearby infrastructure node, after which the communication infrastructure handles message forwarding to the final destination. **Edge and cloud computing infrastructures** provide computational services for drones. One particularly important capability is $offloading(function, dest)$, where a drone requests a remote infrastructure (dest) to execute analytics or data-processing functions. The available analytics functions are defined by the computing infrastructure and may include image analysis, machine-learning inference, mapping, optimization, or other computationally intensive operations. Such services allow resource-constrained drones to leverage powerful external computing resources without carrying the associated hardware.

Several research and engineering tasks are required to realize the service layer. First, services associated with drone-centric devices must be systematically identified and specified. Similarly, services provided by external infrastructures must be characterized and incorporated into the architecture. We will employ LLMs to assist in identifying service categories from existing UAV systems, software platforms, technical documentation, and application case studies. Services will be formally described using the PT-SOA model to facilitate discovery, reasoning, and composition. LLMs will further be used to assist with the generation of service specifications and the extraction of service descriptions from existing documentation and information resources.

## C. The Third Layer: Service Composition and Collaboration Layer (SCC)

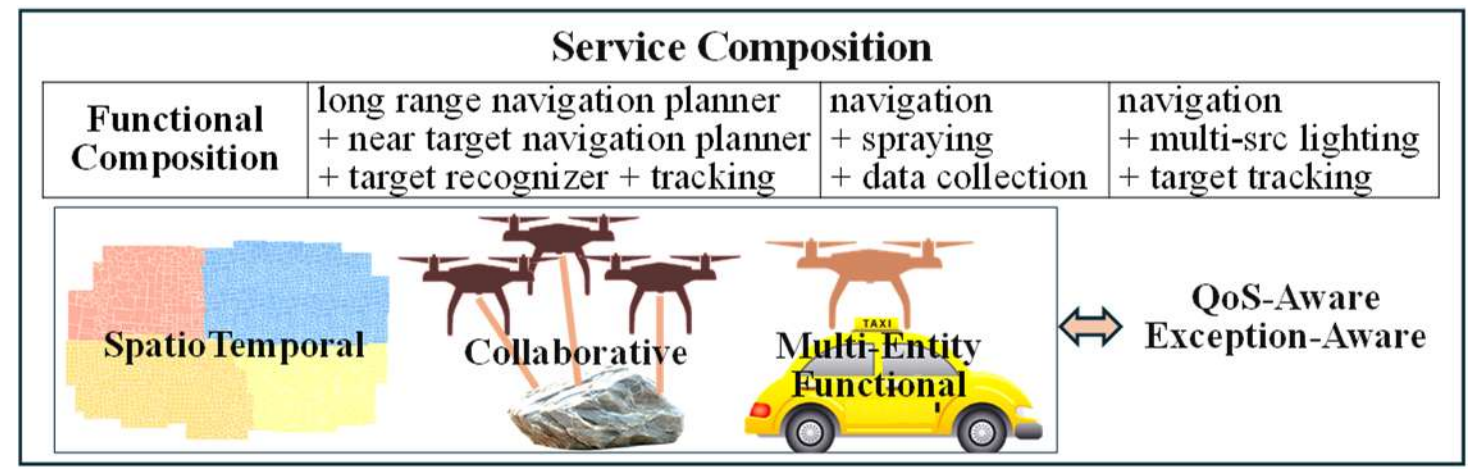

Fig. 3. Examples of some UAV service composition techniques.

Real-world missions require coordinated operations involving multiple drones, various infrastructures, and human operators. In OmniDroneX, we use LLMs and AI planners to build a powerful automated service composition engine for the SCC layer. Although service composition in UAV ecosystems can take diverse forms, we analyze common mission categories and classify the corresponding service composition and coordination techniques accordingly. Some of the categories of compositions techniques are illustrated in Fig. 3.

**Spatiotemporal composition**. Many drone missions focus on covering large geographic areas to search for objects, monitor activities, track targets, or create maps. Examples include smart-city monitoring, post-disaster search and rescue, border surveillance, wildlife monitoring, infrastructure inspection, and environmental mapping. Another class of missions involves providing delivery or transport services across a geographical region or along a planned route. Examples include package delivery, medical supply distribution, debris collection after disasters, etc. Yet another class of mission involve temporary communication-network deployment after disasters or massive communication failures. These missions often require drones to coordinate their movements to maximize spatial coverage with desired overlaps. The situation may also evolve during execution. For example, a surveillance mission may initially monitor a small number of targets, but once additional suspicious activities are detected, more drones may need to be redeployed to the affected area. Similarly, in a disaster assessment mission, newly discovered hotspots may require additional sensing resources. Spatiotemporal composition is the most suitable mechanism to statically or dynamically orchestrate drone services according to spatial and temporal partitions.

**QoS-aware service composition.** In many missions, multiple drones or service providers can perform the same task, but with different performance characteristics. Examples include package delivery services where several drones are available with different battery levels, payload capacities, flight speeds, or operational costs. Similarly, during a surveillance mission, several image-processing services may be available, each offering different levels of accuracy and latency. Mission planners must select the combination of services that best satisfies mission objectives while meeting constraints on cost, reliability, energy consumption, response time, or sensing quality. These requirements lead to QoS-based composition (or service selection), where services are discovered and selected according to quality-of-service criteria. Such composition mechanisms have been extensively investigated in service-oriented computing and multi-agent systems.

**Collaborative composition**. Certain missions involve tasks that exceed the capabilities of a single drone and require close collaborations of multiple drones. Examples include transporting heavy supplies to disaster zones, moving construction materials on job sites, carrying large sensing equipment, or evacuating hazardous objects. Multiple drones may jointly lift and transport a payload whose weight exceeds the capacity of any individual drone. Cooperative inspection missions may also require drones to maintain formations around large structures. These missions require collaborative composition, in which drones coordinate their actions tightly to achieve a common physical objective.

**Functional composition**. Some missions cannot be achieved through a single service function and instead require combining multiple capabilities into a coordinated workflow. Consider a long-range infrastructure inspection mission. Before departure, drones may need weather forecasting services to identify safe flight windows. Ground transportation services may be used to move drones closer to the operational area to conserve energy. During flight, navigation services may interact with airspace management systems to avoid restricted zones. Collected imagery may then be processed by AI-based defect-detection services. Similar compositions arise in disaster response, environmental monitoring, and smart-city operations. Since these missions require combining heterogeneous services to achieve high-level goals, they require functional composition, where service combinations are generated through reasoning and planning. AI planning techniques are particularly useful because they can automatically determine which services should be invoked and in what sequence.

**Holistic composition incorporating exceptions**. Real-world missions rarely proceed exactly as planned. Unexpected events such as battery depletion, adverse weather, communication failures, sensor malfunctions, airspace restrictions, or newly emerging mission priorities may occur during execution. For example, if several drones experience low battery levels while monitoring a wildfire, replacement drones or battery-swap infrastructure may need to be incorporated into the mission plan. If strong winds make part of the operational area inaccessible, tasks may need to be reassigned to more capable drones. During search-and-rescue operations, the discovery of survivors may trigger entirely new transportation and medical-support tasks. Such situations require composition with exceptions, where planning and service composition mechanisms explicitly reason about contingencies and mitigation strategies. Rather than producing a single static plan, the system performs holistic mission planning that incorporates alternative workflows, recovery actions, and adaptive reconfiguration strategies.

**Composite compositions**. A useful observation is that these composition mechanisms are not mutually exclusive. A large-scale disaster response mission may simultaneously require *spatiotemporal composition* for area coverage, *QoS-based composition* for selecting the most suitable resources, *collaborative composition* for heavy-object transportation, *functional composition* for integrating heterogeneous services, and *holistic composition* for handling failures and unexpected events. Together, they form a hierarchy of composition capabilities needed to support increasingly complex autonomous drone missions.

In OmniDroneX, we leverage LLMs and AI planning techniques to implement a composition agent capable of automatically generating executable workflows directly from high-level mission specifications. The composition agent supports both functional composition and QoS-based composition. Specifically, LLMs are used to translate service descriptions in natural language into formal PDDL specifications, including IOPEs of the services, from service documentation, web resources, and manual descriptions. Similarly, missions are described at a high level in natural languages and are translated into PDDL goals, capturing both functional and QoS requirements. When imprecision, incomplete, or inconsistent specifications are detected by a validator, LLMs can work with the user to iteratively refine them until they are well-defined. Then, the planning engine synthesizes executable workflows that orchestrate multiple drones and services to achieve mission objectives efficiently and robustly. A key research task within this framework is to investigate prompt engineering techniques to optimize the interaction between users and LLMs to reach valid mission specifications.

### D. The Fourth Layer, The Marketplace Layer

The fourth layer of OmniDroneX is the marketplace layer, which enables the discovery, sharing, and orchestration of UAV resources and services among diverse stakeholders. This layer builds upon the services and compositions provided by the lower layers to support complex, multi-entity missions. Unlike conventional marketplaces that primarily handle standalone items or simple services, the OmniDroneX marketplace is designed to support composite service offerings, where multiple drones, infrastructures, and coordinated workflows are integrated to accomplish complex mission objectives.

The marketplace involves two primary categories of participants: device owners and users of UAV services. Owners register drones, drone-centric devices, and infrastructure components within the marketplace, specifying metadata such as capabilities, operational constraints, availability schedules, pricing models, and sharing or security policies. The marketplace ensures that participating owners are verified and that their resources can be reliably integrated into mission execution. Users represent entities that wish to accomplish mission objectives through the system. They register functional goals, desired quality-of-service (QoS) requirements, and operational constraints, forming the demand side of the marketplace.

A key feature of the Marketplace Layer is the **LLM-assisted interactive mission specification interface**, which allows users to describe mission requirements in natural language. Instead of requiring structured mission definitions upfront, users can express high-level intent, which is then iteratively refined by an LLM-based assistant into a formal mission specification. During this process, the system may ask clarifying questions and progressively construct a structured task definition that includes mission type, geographic region, resource requirements, and constraints. The LLMs further translate the user intent into multiple executable workflow solutions, each associated with different levels of functionality coverage, quality-of-service attributes, execution time, and cost estimates. This mechanism is conceptually similar to platforms such as Uber or Airbnb, where users are presented with multiple service options and can select based on trade-offs.

The marketplace exposes a $Task\ Registry\ Service$ (TRS), which provides a standardized interface for submitting the finalized mission specification. Once a task is registered, the marketplace invokes the $composition\ agent$ in the SCC layer to generate executable mission workflows. The composition agent identifies suitable drones and infrastructure resources, composes single-drone services into coordinated workflows, and integrates services from external infrastructure. The resulting executable mission plan can be deployed across heterogeneous UAV and infrastructure resources.

Trustworthiness mechanisms are essential in the marketplace to ensure that both owners and users are verified and that their interactions comply with system policies. In OmniDroneX, trustworthiness is established through $multi\text{-}layered\ verification\ processes$ , including identity validation, device certification, and capability verification. Owners' drones and devices can be validated using automated tests and certifications that confirm their capabilities, operational limits, and compliance with safety standards, while users can be verified via identity checks, subscription status, and prior mission history. To maintain ongoing reliability, the marketplace incorporates a $reputation\ and\ rating$ system for both owners and users. After each mission, performance metrics such as timeliness, resource availability, mission success, and adherence to operational constraints are recorded and contribute to a dynamic trust score. Owners with consistently high scores may gain priority in service allocation, while users with reliable behavior can access premium services.

Advanced mechanisms further strengthen trust and accountability of the marketplace. $Blockchain\text{-}based\ audit\ trails$ can record all service registrations, mission executions, and transactions, providing tamper-proof logs that ensure transparency and enable post-hoc verification of claims or disputes. $Smart\ contracts$ can enforce Service-Level Agreements (SLAs) automatically, triggering penalties or notifications if drones or infrastructures fail to meet specified QoS parameters. Additionally, $AI\text{-}driven\ anomaly\ detection$ can analyze mission telemetry, communication logs, and sensor data in real time to identify unusual or potentially unsafe behaviors, such as resource misuse, unauthorized device activation, or performance degradation. These AI agents can either alert human operators or autonomously adjust service allocation to mitigate risks. The overall marketplace design ensures a secure, transparent, and resilient ecosystem, allowing resources to be shared reliably, services to be executed safely, and UAV missions to operate efficiently and robustly under diverse operational and regulatory conditions.

### E. The Fifth Layer: Digital Twin and Validation Layer (DTV)

The DTV layer provides mechanisms for simulation, experimentation, verification, knowledge extraction, and continuous improvement of the OmniDroneX ecosystem. Given the safety-critical nature of many UAV applications, rigorous validation is essential before missions are deployed in real-world environments.

A primary function of this layer is the creation of digital twins representing drones, infrastructures, environments, and mission workflows. Digital twins provide virtual representations of physical resources that enable realistic simulation of system behavior under diverse operational conditions. Such models support mission planning, performance evaluation, failure analysis, and what-if analyses without exposing physical assets to unnecessary risks. Simulations may model drone dynamics, communication performance, battery consumption, environmental conditions, weather impacts, sensing accuracy, infrastructure availability, and collaborative behaviors. Through simulation, users can assess mission feasibility, identify bottlenecks, and compare alternative service compositions before committing physical resources.

OmniDroneX further envisions a knowledge-driven approach to ecosystem evolution. Historical mission reports, operational experiences, scientific publications, technical documentation, and publicly available case studies may be continuously collected and analyzed. LLMs can assist in extracting service primitives, identifying recurring mission patterns, discovering collaboration strategies, and constructing reusable mission templates.

Case-study repositories constitute another important component of this layer. Existing drone applications in domains such as infrastructure inspection, precision agriculture, disaster response, environmental monitoring, logistics, smart cities, and public safety can be organized into structured knowledge bases. These repositories provide valuable resources for understanding operational requirements, identifying reusable capabilities, and validating architectural concepts.

The Digital Twin and Validation Layer may also support synthetic mission generation. LLMs can generate hypothetical mission scenarios, stress-test service compositions, and explore previously unconsidered combinations of resources and capabilities. Such capabilities can accelerate the discovery of new drone services and contribute to the continuous expansion of the OmniDroneX ecosystem. Ultimately, this layer provides the foundation for evidence-driven development and long-term ecosystem evolution.

### F. The Sixth Layer: Execution, Monitoring, and Mitigation Layer (EMM)

The EMM layer in OmniDroneX is responsible for the real-time deployment, supervision, and adaptive control of UAV missions. Once missions are composed and validated in the lower layers, this layer ensures that plans are executed safely, efficiently, and in accordance with the defined QoS and operational constraints. It provides continuous monitoring of drones, payloads, sensors, communication links, and external infrastructures, collecting telemetry and status information to detect deviations from expected behavior. The layer integrates real-time decision-making capabilities to respond to dynamic environmental conditions, unforeseen obstacles, hardware failures, or communication disruptions.

OmniDroneX adopts *predictive maintenance* which analyzes sensor data and operational logs to anticipate component degradation or battery depletion, reducing mission risk and improving resource utilization. Additionally, the EMM layer can implement *dynamic resource reallocation*, enabling drones to adaptively share workloads, reroute through alternative infrastructures, or request support from mobility and battery supply services when necessary. Integration with the Digital Twin and Validation Layer allows continuous feedback loops, where real-time mission performance is compared with simulated expectations, and discrepancies can inform both immediate mitigation and long-term system improvements.

## IV. Challenges and Future Research Directions

While OmniDroneX presents a vision for an integrated, LLM-assisted Drone-as-a-Service ecosystem, realizing it requires significant advances across multiple research domains. We outline several important challenges and future research directions that must be addressed to transform the architecture into a practical platform.

(1) **libUAV for interoperation**. Current drone ecosystems suffer from fragmented APIs and incompatible software stacks. Similar functionalities are often implemented differently across vendors, creating substantial integration efforts for application developers. Identifying a universal set of primitive functions that can effectively capture drone capabilities and facilitate upper-level seamless integration and vendor-agnostic system development is essential. The same needs apply to the service layer where the bi-layered libUAV shall incorporate a universal foundational set of UAV-related services. LLMs may provide promising mechanisms for extracting primitive capabilities and universal services from technical documentation, operational manuals, and mission reports. Research is needed to establish systematic procedures for such LLM-assisted approach and to develop verification and validation methodologies to ensure that such processes yield accurate, consistent, and complete libUAV primitive functions and services definitions. Further research is also needed to employ LLMs for semi-automated development of mapping functions that bind libUAV abstractions to platform-specific APIs by analyzing vendor SDKs, API documentation, and implementation examples. Once such mappings have been established for a representative set of UAV platforms, transfer-learning and knowledge-reuse techniques may enable the automated extension and evolution of libUAV to newly introduced vendor platforms with substantially reduced human effort.

(2) **Formal model and methodology for physical layer composition**. In the physical layer of OmniDroneX architecture, we have identified the power of capability augmentation from physical layer composition. Research is required to develop systematic methodologies to identify composable physical entities for extending the capabilities of UAVs via such composition. Also, it is necessary to develop a formal model for inferencing the new capabilities of the composed entity from existing capabilities of the entities being composed. Currently we attempt to use mobility as the basis for such capability composition, e.g., a sensor's capability is enhanced by drone's mobility. But this may not be applicable for general physical compositions. For example, a drill with attached sponge makes an almost-automated cleaner. Establishing an effective paradigm for modeling and inferencing of physical compositions will be an interesting and critical research direction.

(3) **Formal model for UAV service definitions**. Service specification models and ontologies are needed to specify the functionalities, operational constraints, resource requirements, and QoS properties of UAVs and other physical entities, as well as their services, to facilitate upper level compositions, especially for automated compositions. PT-SOA needs to be adapted to a UAV-centric service model to address such modeling requirements. It is important to note that even with well-designed formal models to guide specification, generating concrete specifications for each device and service at scale can become highly labor-intensive due to the large number of heterogeneous entities and the wide variety of services involved. Such effort is only practically feasible through automation. Therefore, an important challenge is how to leverage LLMs to automatically or semi-automatically generate such specifications, thereby significantly reducing manual effort.

(4) **LLM-assisted mission understanding and planning**. OmniDroneX relies heavily on LLMs to translate user goals into executable mission specifications. While recent advances demonstrate promising capabilities, several challenges remain. First, natural language mission descriptions are often ambiguous, incomplete, and/or inconsistent. LLMs must accurately infer user intent, identify required resources, resolve ambiguities, and generate correct planning representations because errors during this process may lead to mission failures or unsafe behaviors. Future work is needed to develop an interactive procedure to progressively derive complete, consistent, and correct mission specifications in both functional and QoS aspects.

(5) **Autonomous service composition and orchestration**. A central vision of OmniDroneX is the ability to dynamically compose heterogeneous resources into mission-specific executable workflows. This is critical because many UAV-related tasks arise dynamically and resource availability, mission requirements, weather conditions, and communication infrastructures may continuously change. Manual compositions do not fit in such dynamic and reactive situations. Techniques for automated mission plan generation, holistic workflow construction, and mission verification still require further investigation. Hybrid approaches that combine LLM reasoning with symbolic planning, constraint solving, and formal verification may provide promising solutions.

(6) **Multi-agent collaboration and swarm intelligence**. Many UAV applications require cooperation among large numbers of drones and supporting infrastructures. Examples include disaster response, environmental monitoring, large-scale inspection, communication restoration, and collaborative transportation. Future research should investigate reusable collaboration primitives, distributed coordination algorithms, adaptive swarm behaviors, and collaborative decision-making mechanisms. LLMs may assist in identifying collaboration patterns from historical missions and operational experiences; however, methods are required to transform such knowledge into executable coordination strategies. Particularly challenging problems include dynamic collaboration strategy formation, collaboration inference, entity synchronization, conflict resolution, self-stabilization, decentralized decision making, collaborative perception, and fault-tolerant mission execution under uncertain conditions.

(7) **Trust, security, privacy for UAV marketplace and mission execution**. Trustworthiness is essential for any large-scale Drone-as-a-Service ecosystem. Marketplace participants may include resource providers, service consumers, software developers, infrastructure operators, and autonomous agents with varying levels of trust and reliability. Future research should leverage techniques in reputation systems, trust models, secure service execution mechanisms, policy-based access control frameworks, and blockchain infrastructures. Privacy-aware mission execution represents another important research direction, particularly for applications involving surveillance, public spaces, and sensitive information.

(8) **Digital twins and continuous learning**. The proposed Digital Twin and Validation layer in OmniDroneX architecture creates opportunities for continuous ecosystem improvement. Future research should investigate how digital twins can be automatically constructed, maintained, and remain synchronized with physical environments. Another promising research direction is continuous learning from historical missions, operational experiences, simulation outcomes, and publicly available case studies. LLMs may serve as knowledge extraction engines that identify new service primitives, collaboration patterns, mission templates, and case study specifications. Such capabilities could ultimately enable self-evolving drone ecosystems that continuously expand their service offerings and improve mission performance over time.

(9) **Toward a Global Drone-as-a-Service ecosystem**. OmniDroneX envisions a future in which drones, infrastructures, software services, and intelligent agents participate in an open service ecosystem. In such an environment, mission requests may be automatically translated into executable workflows that span multiple organizations, geographic regions, and resource providers. Realizing this ecosystem will require advances not only in drone technologies but also in service computing, artificial intelligence, distributed systems, cyber physical systems, and autonomous agent technologies. We believe OmniDroneX provides an important blueprint for exploring this emerging research direction and establishing the foundations for next-generation Drone-as-a-Service ecosystems.

## V. CONCLUSION

OmniDroneX is a vision for a unified, LLM-native Drone-as-a-Service ecosystem. By integrating physical capability augmentation, service-oriented abstractions, automated composition, and ecosystem-scale coordination, it redefines how drone missions can be easily and comprehensively specified, and how mission plans can be effectively constructed, tested, and executed. A central theme of OmniDroneX is the pervasive use of LLMs to bridge the gap between heterogeneous low-level device primitives and high-level human intent, enabling automation across specification, composition, and adaptation. While significant challenges remain, OmniDroneX provides a conceptual blueprint for future UAV ecosystems that are adaptive, extensible, and continuously evolving. More broadly, it points toward a general paradigm of LLM-assisted cyber-physical systems operating in open, dynamic environments.